\documentclass[aps,pra,preprint,groupedaddress,showpacs,showkeys,nofootinbib,draft]{revtex4-2}
\usepackage{bm,graphicx,amsmath}
\usepackage{centernot}
\begin{document}
\newcommand{\crist}[3]{\ensuremath{\left\{^{\,\,#1}_{#2#3}\right\}}}
\newcommand{\Tr}{\mathrm{Tr}}
\newcommand{\bra}[1]{\ensuremath{\langle #1|} }
\newcommand{\ket}[1]{\ensuremath{|#1\rangle}}
\newcommand{\qo}[1]{``#1''}
\newcommand{\diag}{\mbox{diag}}
\newcommand{\D}{{\cal D}}
\newcommand{\DD}{{\centernot D}}
\title{Weyl-invariant derivation of Dirac equation from scalar tensor fields in curved space-time}
\author{Enrico Santamato}
\email{enrico060248@gmail.com}
\affiliation{Dipartimento di Fisica ``Ettore Pancini'', Universit\`{a} di Napoli ``Federico II'', Italy}
\author{Francesco De Martini}
\email{francesco.demartini2@gmail.com} 
\affiliation{Dipartimento di Fisica dell'Universit\`{a} di Roma
``La Sapienza', Italy} 
\affiliation{Accademia Nazionale dei Lincei, via della Lungara
10, Roma 00165, Italy}
\date{\today}
\begin{abstract}
In this work we present a derivation of Dirac's equation in a
curved space-time starting from a Weyl-invariant action
principle in 4+K dimensions. The Weyl invariance of Dirac's
equation (and of Quantum Mechanics in general) is made possible by observing that the difference between the Weyl and the
Riemann scalar curvatures in a metric space is coincident with
Bohm's Quantum potential. This circumstance allows a completely geometrical formulation of Quantum Mechanics, the Conformal
Quantum Geometrodynamics (CQG), which was proved to be useful,
for example, to clarify some aspects of the quantum paradoxes
and to simplify the demonstration of difficult theorems as the
Spin-Statistics connection. The present work extends our
previous derivation of Dirac's equation from the flat Minkowski space-time to a general curved space-time. Charge and the e.m. fields are introduced by adding extra-coordinates and then gauging the associated group symmetry. The resulting Dirac's
equation yields naturally to the correct gyromagnetic ratio $g_e=2$ for the electron, but differs from the one derived in the Standard Quantum Mechanics (SQM) in two respects. First, the coupling with the space-time Riemann scalar curvature is found to be about 1/4 in the CQG instead of 1/2 as in the SQM and, second, in the CQG result two very small additional terms appear as scalar potentials acting on the particle. One depends on the derivatives of the e.m. field tensor and the other is the scalar Kretschmann term $R_{\mu\nu\rho\sigma}R^{\mu\nu\rho\sigma}$. Both terms, not present in the SQM, become appreciable only at distances of the order of the electron Compton length or less. The Kretschmann term, in particular, is the only one surviving in an external gravitational field obeying Einstein's equations in vacuum. These small differences render the CQG theory confutable by very accurate experiments, at least in principle.
\end{abstract}
\maketitle 
\section{Introduction}\label{sec:introduction}
The Conformal Quantum Geometrodynamics (CQG), provides a
realistic and complete geometric approach to Quantum Mechanics
with no adoption of the quantum wave functions or spinors as the primary object of the theory. The CQG stems on the
identification of the {\it difference} between Weyl's and Riemann's scalar curvatures defined in the same metric space with Bohm's quantum potential~\cite{BohmHiley1995}~\cite{Santamato1984,Santamato1985}. The Riemann differs form the Weyl geometry in the fact that the change $\delta\ell$ of the length of a vector in a parallel transport is not zero but changes according to the affine law $\delta\ell=-2\ell\phi_i(q)dq^i$ where $\phi_i(q)$ is dubbed the Weyl vector in the $N$-D metric space spanned by the coordinates $q^i$ $(i=1,\dots,N)$. A consequence of Weyl's geometry is the requirement of invariance of the theory under local change of the metric calibration~\cite{Weyl1952}. The calibration invariance of physical laws is equivalent to the requirement of invariance under Weyl's gauge transformations
\begin{eqnarray} \label{eq:Weylgauge}
   g_{ij}&\rightarrow&\lambda(q)g_{ij}
   \label{eq:gchange}\nonumber\\
   \phi_i(q)&\rightarrow&\phi_i(q)-\frac{1}{2}\partial_i\ln\lambda(q)
   \label{eq:phichange}
\end{eqnarray}
The CQG allows a formulation of Quantum Mechanics that is invariant under the Weyl gauge transformations~(\ref{eq:Weylgauge}). According to Weyl's geometry the space is curved even if the metric is flat, i.e. Euclidean or Minkowskian. This is enough to originate all quantum effects, including Heisenberg's uncertainty relations~\cite{Santamato1988}, entanglement and nonlocal EPR correlations~\cite{DeMartiniSantamato2012a,DeMartiniSantamato2014,
DeMartiniSantamato2014b,DeMartiniSantamato2015} as well as a simple proof of the spin-statistics connection without assuming commuting and anticommuting operators.~\cite{SantamatoDeMartini2015}.
As with gravity of the standard general relativity, the Weyl curvature potential cannot be shielded by any physical device so that the quantum features are universal and unavoidable.
Like differential geometry, the CQG provides a vector-tensor approach to Quantum Mechanics without having recourse to complex wave functions or spinors as primary objects. Spin 1/2 particles can be included in the CQG as point-like particles carrying a fourleg in space-time~\cite{Santamato2008}. This mechanical model was used to obtain Dirac's equation\footnote{Throughout this work, \qo{Dirac's equation} is used instead of \qo{the square of Dirac's equation}, which is an equivalent second-order, rather than first-order differential equation. The square of Dirac's equation in a curved space-time has been explicitly derived from Dirac's  equation of the SQM in the Appendix.} in the Minkowski space-time~\cite{SantamatoDeMartini2013}. The theory is extended here to a curved space-time which is the natural framework of Weyl's geometry. The spin 1/2 particle is introduced by means of geometric rather than mechanical considerations and requiring gauge invariance under the change of the local basis in space-time. The e.m. field is also introduced geometrically, by adding to the space-time coordinates a fifth coordinate for the U(1) symmetry and requiring invariance under the usual e.m. gauge. These extra coordinates are regarded here as additional hidden variables inaccessible to measurements made in space-time~\cite{DeMartiniSantamato2014b}. The correct gyromagnetic ratio of the electron comes out naturally.\\

The work is divided as follows. In Sec.~\ref{sec:lagrangian} we present a lagrangian formulation of the CQG based on Palatini's variational method. Palatini's approach has the advantage to avoid to postulate Weyl's geometry since the beginning. The resulting field equations are the main equations of the CQG. In Sec.~\ref{sec:metric} we introduce a suitable metric interval invariant under local base change and e.m. gauge. In Sec.~\ref{sec:dirac} we briefly discuss the Dirac equation deduced from the field equations presented in Sec.~\ref{sec:lagrangian}. Finally, in Sec.~\ref{sec:conclusions} we draw our conclusions. In the Appendix we present, for the sake of comparison, the square of Dirac's equation in a curved space-time as calculated in the framework of the SQM. \\

\section{The Lagrangian of the CQG}\label{sec:lagrangian}
Let us consider a $N$-D metric space $V_N$ spanned by coordinates $q^i$ ($i=1,\dots,N$) with metric tensor $g_{ij}(q)$ and torsion free affine connections $\Gamma^i_{jk}(q)=\Gamma^i_{kj}(q)$. In the space $V_N$ we define the action integral
\begin{equation}\label{eq:action}
   I = \int d^nq\sqrt{|g|}[\rho(g^{ij}\partial_i\sigma\partial_j\sigma + R)]
\end{equation}
where $\sigma(q)$ is a scalar field and $R$ is the scalar curvature built from the connections $\Gamma^i_{jk}$\footnote{The curvature tensor is here defined as $R^i_{jkl}=\partial_k\Gamma^i_{jl}- \partial_l\Gamma^i_{jk}+ \Gamma^i_{sk}\Gamma^s_{jl}-\Gamma^i_{sl}\Gamma^s_{jk}$.} and consider the field equations associated with the independent variations of  $\Gamma^i_{jk}$, $\sigma$, $\rho$. The respective field equations are
\begin{equation}\label{eq:Gammaijk}
  \Gamma^i_{jk}=\crist{i}{j}{k}+
      (\delta^i_j\phi_k+\delta^i_k\phi_j-g_{jk}\phi^i)
\end{equation}
where $\crist{i}{j}{k}$ are the Christoffel symbols calculated
from the metric tensor $g_{ij}$ and $\phi_i$ is given by
\begin{equation}\label{eq:phii}
\phi_i=\frac{1}{N-2}\partial_i\ln\rho
\end{equation}
\begin{equation}\label{eq:cont}
   \partial_i(\sqrt{|g|}\rho g^{ij}\partial_j\sigma) = 0.
\end{equation}
\begin{equation}\label{eq:HJ}
   g^{ij}\partial_i\sigma\partial_j\sigma + R = 0
\end{equation}
We observe that the connections (\ref{eq:Gammaijk}) are Weyl's connections and $R$ is therefore Weyl's scalar curvature. Moreover, Eq.~(\ref{eq:phii}) shows that the  action~(\ref{eq:action}) and the field $\rho$ fix the Weyl gauge to be integrable. The Weyl gauge fixed by $\rho$ will be called the Weyl-$\rho$-gauge. In the following we will use this gauge until specifically mentioned. The Weyl scalar curvature in the Weyl-$\rho$-gauge is given by
\begin{equation}\label{eq:R}
   R = \bar{R}- \gamma^{-2}\left(\frac{\nabla_k\nabla^k\sqrt{\rho}}{\sqrt{\rho}}\right)
\end{equation}
where $\bar R$ is the Riemann scalar curvature built form the Christoffel symbols of the metric $g_{ij}$ and $\gamma$ is a constant given by
\begin{equation}\label{eq:gamma}
   \gamma = \sqrt{\frac{N-2}{4(N-1)}}
\end{equation}
The last term on the right of Eq.~(\ref{eq:R}) is proportional to Bohm's quantum potential, as said before. In this way the connection between Weyl's geometry and Quantum Mechanics is established. In fact, if we introduce the complex field $\Psi$ defined as
\begin{equation}\label{eq:Psidef}
   \Psi = \sqrt{\rho}e^{\frac{\sigma}{\gamma}}
\end{equation}
Eqs.~(\ref{eq:cont}) and (\ref{eq:HJ}) yield the linear wave equation of the CQG, viz.
\begin{equation}\label{eq:confwavequ}
  \nabla_k\nabla^k\Psi+\gamma^2\bar R\Psi= 0
\end{equation}
We may also vary the action~(\ref{eq:action}) with respect of the metric tensor $g_{ij}$. The resulting field equations can be cast in the Einstein's form
\begin{equation}\label{eq:Einstein}
   \bar R_{ij}-\frac{1}{2}g_{ij}\bar R + T_{ij} = 0
\end{equation}
with $T_{ij}$ given by
\begin{eqnarray}\label{eq:Tij}
   T_{ij} &=& \nabla_i\sigma\nabla_j\sigma -\frac{1}{2}g_{ij}\nabla_k\nabla^k\sigma + \nonumber \\ && +\frac{1}{\gamma^2\rho^2}\left(\nabla_i\rho\nabla_j\rho - \frac{1}{2}g_{ij}\nabla_k\nabla^k\rho\right) -
   \frac{1}{\rho}\left(\nabla_i\nabla_j\rho-g_{ij}\nabla_k\nabla^k\rho\right)
\end{eqnarray}
A straightforward calculation shows that Eqs.~(\ref{eq:cont}) and (\ref{eq:HJ}) and, hence, the wave equation (\ref{eq:confwavequ}) are consequences of the field Eqs.~(\ref{eq:Einstein}).\\

In the following we are looking for solutions of the field Eqs.~(\ref{eq:cont}) and (\ref{eq:HJ}) and, hence, solutions of the wave equation (\ref{eq:confwavequ}) describing a single Dirac's electron. Therefore the solutions of Eqs.~(\ref{eq:Einstein}) would eventually describe the metric tensor $g_{ij}$ created by one Dirac electron. Here we assume, instead, that the metric $g_{ij}$ is due to the presence of external bodies of very large mass so that the tensor $T_{ij}$ in Eqs.~(\ref{eq:Einstein}) is negligible with respect to the overwhelming contribution [omitted in Eqs.~(\ref{eq:Einstein})] due external matter. In other word, we consider the electron subjected to an external gravitational field created by other bodies. Here we just limit to observe that the trivial solution $\sigma=$const, $\rho=$const. (i.e. $\Psi=$const.) of our field equations yields $T_{ij}=0$ and Eqs.~(\ref{eq:Einstein}) reduce to Einstein's equations in vacuum.

As final point, we notice that if the field $\rho=\Psi^*\Psi$ has Weyl type $w(\rho)=-(N-2)/2$, the action~(\ref{eq:action}) is Weyl invariant\footnote{A tensor $T$ is said to have Weyl's type $w(T)$ if under the Weyl gauge transformations~(\ref{eq:Weylgauge}) $T$ transforms as $T\rightarrow\lambda^{w(T)}T$. For example $w(g_{ij})=1$, $w(g^{ij})=-1$, $w(\sqrt{|g|})=N/2$, $w(\Gamma^i_{jk})=0$, $w(R)=-1$.}. Hence, although not explicitly, the field equations~(\ref{eq:cont}), (\ref{eq:HJ}), (\ref{eq:confwavequ}) and (\ref{eq:Einstein}) are Weyl invariant. Because $\rho$ depends on the Weyl gauge, its physical interpretation depends on the chosen gauge. In the Weyl-$\rho$-gauge, the field $\rho=\Psi^*\Psi$ can be physically interpreted as a probability density~\cite{Santamato1984a,Santamato1985}, but this interpretations is impossible in other gauges. For example, if we choose a new gauge so that $\rho\rightarrow\rho'=1$, the field $\rho$ of the old $\rho$-gauge is now intimately bound to the new metric $g'_{ij}=\rho^{2/(N-2)}g_{ij}$. We see therefore that in the CQG $\Psi^*\Psi$ has a geometric rather that statistical meaning. Its role is to produce a curvature even in a space which is flat in the Riemann sense. And this curvature produces all quantum effects.
\section{The metric tensor}\label{sec:metric}
In this section we fix the metric tensor $g_{ij}$. We follow the main lines of the well known Kaluza-Klein (KK) gauge theories~\cite{OverduinWesson2004,SalamStrathdee1982,Bailin1987}. To the four coordinates $x^\mu$ ($\mu=0,1,2,3$) of the space-time $M_4$ we add the other $K$ coordinates of the variety of a suitable group $G$ so to have a metric space $V_N = M_4\times G$ of $N=4+K$ dimensions. The $K$ additional coordinates are here the six parameters $y^\alpha$ ($\alpha=1,...,6$) of the proper orthocronous Lorentz group $L_+$ to introduce the quantum spin\footnote{As will be clarified in Sec.~\ref{sec:dirac}, our group $G$ is he quotient group $L_+/R_\zeta$, where $R_\zeta$ is the abelian subgroup of space rotation around the fixed axis $\zeta$.}, and the single parameter $z$ of the U(1) group to introduce the electron charge. Our space has therefore eleven dimensions and coordinates $q^i=\{x^\mu,y^\alpha,z\}$. In $V_{11}$ we consider the eleven independent forms $d\bar s^A=e^A_{\ i}(q)dq^i=\{d\bar s^m,d\bar s^a,d\bar z\}$ ($A=0,...,10$), ($m=0,...,3$), ($a=1,...6$) given by
\begin{eqnarray}\label{eq:dsbar}
   d\bar{s}^m &=& ds^m = e^m_{\ \mu}(x)dx^\mu  \nonumber \\
   d\bar{s}^{mn} &=& \lambda_L\left(ds^{mn} + \omega^{mn}_{\quad p}(x)ds^p\right) \\
   d\bar{z} &=& \lambda_0 dz + \lambda_e A_m(x)ds^m + \lambda_e\lambda_L F_{mn}(x)d\bar{s}^{mn}   \nonumber
\end{eqnarray}
where $\lambda_0$, $\lambda_e$, and $\lambda_L$ are constants having the dimensions of a length. Their are needed to have all forms $d\bar s^A$ with the dimensions of a length.
The four forms $d\bar{s}^m=e^m_{\ \mu}(x)dx^\mu$ are forms in the space-time coordinate $x^\mu$ only and are a pseudo-orthogonal basis in the space-time $M_4$. Therefore we have $\eta_{mn}e^m_{\ \mu}(x)e^n_{\ \nu}(x)= g_{\mu\nu}(x)$, where $\eta_{mn}=\diag{(-1,1,1,1)}$ is the Minkowski metric tensor and $g_{\mu\nu}(x)$ the metric tensor of $M_4$. Then, by construction, a change of the space-time basis $d\bar s^m$ corresponds to a Lorentz transformation, viz. $d\bar{s}^m\rightarrow d\bar{s}'^m=\Lambda^m_{\ n}(x)d\bar{s}^n$, with $\Lambda^m_{\ n}\in L_+$. Moreover, in place of the six forms $d\bar{s}^a$ we prefer to use in Eqs.~(\ref{eq:dsbar}) the six independent components of the $4\times 4$ antisymmetric matrix $d\bar{s}^{mn}$. The advantage is that in a change of the space-time basis $d\bar s^m$ they transform as an antisymmetric Lorentz tensor under the particular coordinate change
\begin{equation}\label{eq:ygauge}
   D(y)\rightarrow D(y')=D(\bar y(x))D(y)
\end{equation}
where $\bar y^\alpha(x)$ are arbitrary functions of $x^\mu$. Finally, the form $d\bar{s}$ is chosen so to be invariant with respect to the particular coordinate change
\begin{equation}\label{eq:zgauge}
   z\rightarrow z' = z + \chi(x).
\end{equation}
where $\chi(x)$ is an arbitrary function of $x^\mu$.\\
The fields $\omega^{mn}_{\quad p}(x)$ and $A_m(x)$ are the gauge fields needed to have the required covariance and invariance under the coordinate changes (\ref{eq:ygauge}) and (\ref{eq:zgauge}), respectively. The gauge fields $\omega^{mn}_{\quad p}(x)$ are the components of the so-called spin connections $\omega^{mn}_{\quad \mu}(x)=-\eta^{np}e^\nu_{\ p}\nabla_\mu e^m_{\ \nu}$ and the gauge fields $A_m(x)$ are the components of the electromagnetic potentials $A_\mu(x)$ in the space-time basis $d\bar s^m$. The gauge invariant fields $F_{mn}(x)$ are the components in the space-time basis of electromagnetic fields $F_{\mu\nu}(x)=\partial_\mu A_\nu - \partial_\nu A_\mu$. Finally, the antisymmetric tensor $ds^{mn}$ in Eq.~(\ref{eq:ygauge}) is obtained from the Mauer-Cartan equation of the Lorentz group, viz.
\begin{equation}\label{eq:MauerCartan}
   dD(y) D^{-1}(y) = J_ads^a = -\frac{1}{2} J_{mn}ds^{mn}
\end{equation}
where $J_{mn}$ is the antisymmetric Lorentz tensor of the six generators $J_a=\{\bm J,\bm K\}$ for rotation and boosts in the same representations, respectively. In spite of the notation, in a change of the basis $d\bar s^m\rightarrow d\bar s'^m$ in space-time, the forms $ds^{mn}$ defined in Eq.~(\ref{eq:MauerCartan}) transforms inhomogeneously and are not the components of a contravariant Lorentz tensor. The same occurs, however, for the spin connections forms $d\omega^{mn}=\omega^{mn}_{\quad p}(x)ds^p$ so that the inhomogeneous terms compensate and the the forms $d\bar{s}^{mn}$ in Eqs.~(\ref{eq:dsbar}) do transform as the components of an antisymmetric Lorentz tensor as said above.\\
The meaning of the coordinate change (\ref{eq:ygauge}) can be described by the following geometric picture. The four forms $d\bar s^m$ in Eqs.~(\ref{eq:dsbar}) define a local reference frame (vierbein) at any point $x^\mu$ of spaace-time. Let us consider at some given point $Q(x)$ a pseudo-orthonormal frame $dS^m$. Then the six coordinates $y^\alpha$ can be considered as the six Euler angles fixing the orientation of $dS^m$ with respect to the reference basis $d\bar s^m$ located at $Q(x)$. In general, we may take $d\bar s^m=\Lambda(y)^m_{\ n}dS^n$, where $\Lambda(y)\in L_+$ is a Lorentz matrix. Let us consider now the change of the reference basis $d\bar s^m\rightarrow d\bar s'^m = \Lambda(\bar y(x))^m_{\ n}d\bar s_n$ with arbitrary parameters $\bar y^\alpha(x)$. We have, in the new basis, $d\bar s'^m=\Lambda(y')^m_{\ n}dS^n$ with different set of angles $y'^\alpha$. These relations all together imply Eq.~(\ref{eq:ygauge}) with $D(y) = \Lambda(y)^m_{\ n} \in L_+$.\\
Associated with the contravariant forms $d\bar s^A$ in $V_{11}$ are the covariant directional derivatives
\begin{eqnarray}\label{eq:dirder}
   \frac{\partial}{\partial \bar s^m} &=&
        \frac{\partial}{\partial s^m}-
        \lambda_e A_m(x)\frac{\partial}{\lambda_0\partial z}-
        \frac{1}{2}\omega^{pq}_{\quad m}(x)\frac{\partial}{\partial s^{pq}} \nonumber \\
   \frac{\partial}{\partial\bar s^{mn}} &=&
        \frac{\partial}{\lambda_L\partial s^{mn}}-
        \lambda_e\lambda_LF_{mn}(x)\frac{\partial}{\lambda_0\partial z}  \\
   \frac{\partial}{\partial\bar s} &=& \frac{\partial}{\lambda_0\partial z}  \nonumber
\end{eqnarray}
Finally, together with the forms $d\bar s^A$ we introduce in $V_{11}$ the gauge invariant metric tensor $g_{ij}$ as
\begin{equation}\label{eq:ds2}
   ds^2 = g_{ij}(q)dq^idq^j=\eta_{AB}d\bar s^Ad\bar s^B=\eta_{mn}d\bar s^md\bar s^n-\frac{1}{2}d\bar s_{mn}d\bar s^{mn}+d\bar z^2
\end{equation}
By construction, $\eta_{AB}$ is a diagonal matrix with entries $\pm 1$ so that the metric $g_{ij}(q)$ induces in $V_{11}$ a Riemann geometry.  The scalar Riemann curvature $\bar R$ induced in $V_{11}$ by the metric (\ref{eq:ds2}) is given by
\begin{eqnarray}\label{eq:Rbar}
   \bar R &=& R_4+\frac{6}{\lambda_L^2}-\frac{3\lambda_e^2}{4}F_{\mu\nu} F^{\mu\nu}-\frac{\lambda_L^2}{8}R_{\mu\nu\rho\sigma}
   R^{\mu\nu\rho\sigma} + \nonumber \\
   &{}+& \frac{\lambda_e^2\lambda_L^2}{8}\left[F^{\mu\nu}F^{\rho\sigma}
   (2R_{\mu\nu\rho\sigma}-\lambda_L^2R_{\mu\nu}^{\quad\kappa\lambda}
   R_{\rho\sigma\kappa\lambda})-2(\nabla_\rho F_{\mu\nu})(\nabla^\rho F^{\mu\nu})\right]
\end{eqnarray}
where $R_{\mu\nu\rho\sigma}=R_{\mu\nu\rho\sigma}(x)$ is the Riemann tensor and  $R_4=R_4(x)$ is the Riemann scalar curvature induced in the space-time by the metric $g_{\mu\nu}(x)$. In Eq.~(\ref{eq:Rbar}) we passed to the absolute coordinates $F_{\mu\nu}=F_{mn}e^m_{\ \mu}e^n_{\ \nu}$ and $R_{\mu\nu\rho\sigma}=R_{mnrs}e^m_{\ \mu} e^n_{\ \nu} e^r_{\ \rho} e^s_{\ \sigma}$.
\section{The Dirac equation}\label{sec:dirac}
Having defined the metric tensor $g_{ij}$ in Eq.~(\ref{eq:ds2}) we may now write down explicitly the conformal wave equation~(\ref{eq:confwavequ}). After having expressed the Laplace-Beltrami operator $\nabla_k\nabla^k$ using Eqs.~(\ref{eq:dirder}) we obtain
\begin{eqnarray}\label{eq:wavequ}
   &&g^{\mu\nu}\left(\nabla_\mu-\frac{1}{2}\omega^{pq}_{\quad\mu}(x)\frac{\partial}{\partial s^{pq}}-\lambda_e A_\mu(x)\frac{\partial}{\lambda_0\partial z}\right)\left(\nabla_\nu-\frac{1}{2}\omega^{pq}_{\quad\nu}(x)\frac{\partial}{\partial s^{pq}}-\lambda_e A_\nu(x)\frac{\partial}{\lambda_0 \partial z}\right)\Psi+ \nonumber \\
   {}&+&\frac{1}{2}\eta^{mp}\eta^{nq}\left(\frac{\partial}{\lambda_L\partial s^{mn}}-\lambda_e\lambda_L F_{mn}(x)\frac{\partial}{\lambda_0\partial z}\right)\left(\frac{\partial}{\lambda_L\partial s^{pq}}-\lambda_e\lambda_L F_{pq}(x)\frac{\partial}{\lambda_0\partial z}\right)\Psi+ \\{} &+& \frac{1}{\lambda_0^2}\frac{\partial^2\Psi}{\partial z^2}-\gamma^2\bar R \Psi = 0  \nonumber
\end{eqnarray}
with $\bar R$ given by Eq.~(\ref{eq:Rbar}). We now pose
\begin{eqnarray}\label{eq:const}
   \lambda_e &=& e \lambda_0 \\ \nonumber
   \lambda_L &=& \sqrt{\frac{3}{2}}\gamma\lambda_0
\end{eqnarray}
with $\gamma=\frac{3}{2\sqrt{10}}$ given by Eq.~(\ref{eq:gamma}) for $N = 11$ and $e$ the electron charge. The length $\lambda_L$ in Eqs.~(\ref{eq:const}) is chosen so to cancel the term proportional to $F_{\mu\nu}F^{\mu\nu}=F_{mn}F^{mn}$ in Eq.~(\ref{eq:wavequ}). From Eq.~(\ref{eq:const}) we see that $\lambda_e$ and $\lambda_L$ are determined once $\lambda_0$ is given. As we shall se $\lambda_0$ turns out to be of the order of the electron Compton length $\lambda_C$ [see E.~(\ref{eq:lambda0}) below]. This choice avoids the huge masses which plague the KK theories~\cite{Bailin1987}. Insertion of the constants (\ref{eq:const}) into Eqs.~(\ref{eq:wavequ}) and (\ref{eq:Rbar}) yields
\begin{eqnarray}\label{eq:wavequfinal}
   &&g^{\mu\nu}\left(\nabla_\mu-\frac{1}{2}\omega^{pq}_{\quad\mu}(x)\frac{\partial}{\partial s^{pq}}-e A_\mu(x)\frac{\partial}{\partial z}\right)\left(\nabla_\nu-\frac{1}{2}\omega^{pq}_{\quad\nu}(x)\frac{\partial}{\partial s^{pq}}-e A_\nu(x)\frac{\partial}{\partial z}\right)\Psi- \nonumber \\
   {}&-&e\eta^{mp}\eta^{nq}F_{mn}(x)\frac{\partial^2\Psi}{\partial s^{pq}\partial z}+\frac{1}{\lambda_0^2}\left(\frac{\partial^2\Psi}{\partial z^2}+\frac{1}{3\gamma^2}\eta^{mp}\eta^{nq}\frac{\partial^2\Psi}{\partial s^{mp}\partial s^{pq}}\right)+\\
   {}&+&\frac{3\gamma^2 e^2 \lambda_0^2}{4}F_{\mu\nu}F^{\mu\nu}\left(\frac{\partial^2}{\partial z^2}+1\right)\Psi-\frac{4\Psi}{\lambda_0^2}-  \nonumber \\
   {}&-&\left(\gamma^2 R_4 -\frac{3\gamma^4\lambda_0^2}{16}R_{\mu\nu\rho\sigma}R^{\mu\nu\rho\sigma}+
   \frac{3\gamma^4e^2\lambda_0^4}{32}X(x)\right)\Psi = 0  \nonumber
\end{eqnarray}
with $X(x)$ given by
\begin{equation}\label{eq:X}
   X(x)=F^{\mu\nu}F^{\rho\sigma}
   (4R_{\mu\nu\rho\sigma}-3\gamma^2\lambda_0^2R_{\mu\nu}^{\quad\kappa\lambda}
   R_{\rho\sigma\kappa\lambda})-4(\nabla_\rho F_{\mu\nu})(\nabla^\rho F^{\mu\nu})
\end{equation}
As made in Ref.~\cite{SantamatoDeMartini2013} we take the solution $\Psi$ of Eq.~(\ref{eq:wavequfinal}) as a harmonic expansion with respect to the $D^{(0,1/2)}\oplus D^{(1/2,0)}$ representation of the factor group $(L_+/R_\zeta)\times U(1)$, where $R_\zeta$ is the abelian subgroup of the rotations around the space $\zeta$ axis of the fixed fourleg $dS^m$ at the space-time point $Q(x)$. The choice of the quotient group $L_+/R_\zeta$ is due to the conservation of the intrinsic helicity $s=1/2$ of the electron~\cite{SantamatoDeMartini2013}\footnote{The conserved intrinsic helicity plays a crucial role in the proof of the spin-statistics connection~\cite{SantamatoDeMartini2015}.}. We pose, therefore,
\begin{equation}\label{eq:Psiexpansion}
   \Psi^{(s)}(x,y,z)=\left[\left(D^{(0,1/2)}[\Lambda^{-1}(y)]\right)^{(s)}\Psi_R(x) +\left(D^{(1/2,0)}[\Lambda^{-1}(y)]\right)^{(s)}\Psi_L(x)\right]e^{iz}
\end{equation}
where $\Lambda\in L_+$, the label $s=1/2$ is the intrinsic helicity and $D^{(\cdot,\cdot)}[\Lambda(y)]^{(s)}$ is the first row of the corresponding $2\times 2$ matrix~\cite{SantamatoDeMartini2015}. The expansion coefficients $\Psi_R(x)$ and $\Psi_L(x)$ are identified, respectively, as the right and left two-component spinors of Dirac's electron in the Weyl chiral representation. In view of the Mauer-Cartan Eq.~(\ref{eq:MauerCartan}), we can easily see that the action of the differential operator $-i\frac{\partial}{\partial s^{mn}}$ applied to $\Psi^{(s)}$ is equivalent to the action of the Quantum Mechanical operator of the Lorentz group generators in the (0,1/2) representation
\begin{equation}\label{eq:Jmn}
J_{mn}=\left(
    \begin{matrix}
          0 & - K_a  \\
          K_a & J_{ab}
    \end{matrix}
       \right)
\end{equation}
with $K_a = (i/2)\sigma_a$ and $J_{ab}= (1/2)\epsilon_{ab}^{\quad c}\sigma_c$ [$J_{12}=1/2\sigma_3$. etc.] with Pauli matrices $\sigma_a$ $(a,b,c=1,2,3)$. Analogously, the differential operator $-i\partial/\partial z$ is equivalent to the action of the Quantum Mechanical generator of the U(1) group.\\
Therefore, insertion of Eq.~(\ref{eq:Psiexpansion}) in to Eq.~(\ref{eq:wavequfinal}) together with Eq.~(\ref{eq:X}) yields
\begin{eqnarray}\label{eq:Dirac2}
   &&g^{\mu\nu}\left[\left(-i\nabla_\mu-\frac{1}{2}\omega^{pq}_{\quad \mu}(x) J_{pq}-e A_\mu(x)\right)\left(-i\nabla_\nu-\frac{1}{2} \omega^{pq}_{\quad \nu}(x) J_{pq}-e A_\nu(x)\right)\right]\Psi_D - \nonumber\\
   &{}-& e \left(\bm H(x)\cdot\bm\Sigma-i\bm E(x)\cdot\bm\alpha\right)\Psi_D+\left(m_e^2+\gamma^2R_4\right) \Psi_D=V(x)\Psi_D
\end{eqnarray}
where $\Psi_D=\left(\begin{smallmatrix} \Psi_R \\ \Psi_L \end{smallmatrix}\right)$ is the four component Dirac spinor, $\bm\Sigma=\left(\begin{smallmatrix}
        \bm\sigma & 0 \\
        0 & \bm\sigma
        \end{smallmatrix}\right)$,
$\bm\alpha=\left(\begin{smallmatrix}
        \bm\sigma & 0 \\
        0 & -\bm\sigma
        \end{smallmatrix}\right)$,
and
\begin{eqnarray}\label{eq:V}
   V(x)&=&\frac{3\left(1+4\gamma^2\right)\lambda_C^2}{32}
   \left\{2R_{\mu\nu\rho\sigma} R^{\mu\nu\rho\sigma}-\left(1+4\gamma^2\right)e^2\lambda_C^2\times\right. \nonumber \\ &&\left.\times\left[F^{\mu\nu}F^{\rho\sigma}\left( 4R_{\mu\nu\rho\sigma}-3\left(1+4\gamma^2\right)\lambda_C^2R_{\mu\nu}^{\quad\kappa\lambda} R_{\mu\nu\kappa\lambda}\right)-4\left(\nabla_\rho F_{\mu\nu}\right)\left(\nabla^\rho F^{\mu\nu}\right)\right]\right\}
\end{eqnarray}
In Eqs.~(\ref{eq:Dirac2}) and (\ref{eq:V}) we have used the identities $1/2F^{mn}J_{mn}=1/2(\bm M\cdot\bm\Sigma-i\bm E\cdot\bm\alpha)$, $1/2J_{mn}J^{mn}=J^2-K^2=3/2$ and finally posed
\begin{equation}\label{eq:lambda0}
   \lambda_0=\frac{\sqrt{1+4\gamma^2}}{\gamma m_e}=\frac{\lambda_C\sqrt{1+4\gamma^2}}{\gamma}
\end{equation}
with $\gamma=\frac{3}{2\sqrt{10}}$. In the equations above we used units with $\hbar=c=1$ so that the electron mass $m_e$ is related to the electron Compton length $\lambda_C$ by $m_e=1/\lambda_C$. The passage to {\it cgs} units in Eq.~(\ref{eq:Dirac2}) and (\ref{eq:V}) is made by the replacements $A_\mu\rightarrow A_\mu/\sqrt{\hbar c}$, with $A_\mu=(\Phi,-\bm A)$, $(\bm E, \bm H)\rightarrow (\bm E,\bm H)\sqrt{\hbar c}$, $e\rightarrow e/\sqrt{\hbar c}$, $\omega^{pq}_{\quad\mu}\rightarrow \omega^{pq}_{\quad\mu}/\hbar$, $\lambda_C\rightarrow \hbar/(m_ec)$.\\
As shown in Eq.~(\ref{eq:V}), the right hand term in Eq.~(\ref{eq:Dirac2}) is proportional to the square of the electron Compton length and is usually negligible. If we neglect the right term of Eq.~(\ref{eq:Dirac2}), we are left with the square of Dirac's equation in a curved space-time with the only difference that the coupling constant with the space-time scalar curvature is here $\gamma^2\simeq 1/4$ instead of $1/2$, which is the result of the quantum calculation [see the Appendix]. The coupling with $\gamma^2\simeq 1/4$ [see Eq.~(\ref{eq:gamma})] is necessary, however, to have the conformal Weyl invariance of Eq.~(\ref{eq:confwavequ}). In the case of the Minkowski space-time, our Eq.~(\ref{eq:Dirac2}) (with $V(x)$ neglected), is precisely the square of Dirac's equation with the right $g$-factor $g_e=2$ for the electron~\cite{landau4}. As it is well known, the square of Dirac's equation is a second order differential equation. Therefore, in addition to all solutions of the first-order Dirac's equation, Eq.~(\ref{eq:Dirac2}) has extra solutions corresponding to a negative mass. However, to isolate the positive mass solutions is straightforward~\cite{landau4}.\\
We conclude this section with a few words about the potential (\ref{eq:V}) present in our Eq.~(\ref{eq:Dirac2}) and absent in the quantum mechanical case [see Appendix]. In the vacuum we have $F_{\mu\nu}=0$, $R_4=0$ and the term proportional to the Kretschmann scalar $K(x)=R_{\mu\nu\rho\sigma}R^{\mu\nu\rho\sigma}$ only survives in Eq.~(\ref{eq:Dirac2}) and (\ref{eq:V}). The Kretschmann scalar is everywhere negligible except, for example, near the singularity at $r=0$ of the Schwarzschild gravitational metric created by a non rotating spherical body of mass $M$. In this case, $K(x)=12r_S^2/r^6$ where $r_S=2GM/c^2$ is the Schwarzschild radius. To estimate of the order of magnitude of the $K$-term in Eq.~(\ref{eq:Dirac2}) we compare $\lambda_C^2 K(x)$ with the rest mass term $m_e^2=1/\lambda_C^2$. We see that the $K$-term is negligible when $r\gg\lambda_C^{2/3}r_S^{1/3}$. Now, to observe the effects of $K(x)$ on the electron, its position $r$ must be out of the horizon of the events of the mass $M$, i.e. $r\gg r_S$. On the other hand, because the uncertainty on the electron position is about $\lambda_C$ we require that the distance $r$ of the electron from the gravitational essential singularity at $r=0$ should be $r\gg\lambda_C$. These two requirements imply $r\gg \lambda_C^{2/3}r_S^{1/3}$ and the $K$-term in Eq.~(\ref{eq:Dirac2}) can be neglected. In the presence of the electromagnetic field, the dominant term in $V(x)$ is the one proportional to $\nabla_\rho F_{\mu\nu}\nabla^\rho F^{\mu\nu}$. This term is the only one surviving in the Minkowski space-time and is sizeable only if the e.m. fields vary appreciably over the electron Compton length $\lambda_C$. For example, we may consider the electrostatic $E=e/r^2$ field generated by a charge $e$ acting on the electron at distance $r$. Neglecting factors of the order of unit, we compare, as before, $e^4\lambda_C^4/r^6$ with the electron rest mass $1/\lambda_C^2$. We may then neglect the last term in $V(x)$ if $e^4(\lambda_C/r)^6\ll 1$ which implies $r\gg 0.03\lambda_C$, a condition always met in all practical cases.
\section{Conclusions}\label{sec:conclusions}
We derived the square of Dirac's equation from a Weyl invariant Lagrangian in a curved space of $N=4+K$ dimensions. Because Dirac's equation in a curved space assumes that the metric tensor and the e.m potentials are arbitrarily prescribed, we considered only the field equations of the fields $\rho$ and $\sigma$ associated with the particle and disregarded all other field equations deduced from our Lagrangian. The nonlinear field equations for $\rho$ and $\sigma$ can be gathered in a {\it linear} field equation for the single complex field $\Psi$. This simplification was made possible because the difference between the Weyl and the Riemann scalar curvatures is equal to the quantum potential of Bohm's approach to quantum mechanics. This identification, however, requires a curved space endowed with Weyl's geometry as considered in this work. This approach, dubbed CQG, was proved to be useful in particular cases as, for example, to gain an easy proof of the spin-statistics theorem in both relativistic and non-relativistic quantum mechanics. All equations of the CQG are Weyl-gauge invariant, but the physical interpretation may depend on the chosen Weyl gauge. For example, the statistical interpretation of $\rho=\Psi\Psi^*$ as describing a particle moving in a space with given metric tensor $g_{ij}$ is possible only in the gauge fixed by our Lagrangian. In other gauges the statistical interpretation is impossible because $\rho$ is intimately bound to the metric tensor. To derive Dirac's equation we introduced e.m. gauge fields by exploiting the standard methods of Kaluza-Klein theories in $4+K$ dimensions. From the quantum point of view, the extra-coordinates appear as non local hidden variables of the theory and are inaccessible to measurements made in the 4D space-time. As said above, the CQG yields the square of Dirac equation. This result was compared with the square of Dirac equation derived from the SQM. Both approaches show a coupling of Dirac's four component spinor with the space-time Riemann scalar curvature, acting as a scalar potential. The coupling constant, however, was $1/2$ in the Quantum Mechanical approach and about $1/4$ in the CQG approach. Moreover, the CQG derivation has a further scalar potential $V$ quadratic in the space-time curvature tensor and in the space-time derivatives of the e.m. field tensor. These terms have been proved to be negligible in all practical cases. These very small differences between the CQG and the standard Quantum Mechanics could be tested, in principle, by very accurate experiments. The extension of the present work to Yang-Mills gauge potentials is straightforward but the introduction of a further Higgs' field seems unavoidable to provide their mass.
\section{Appendix}\label.
Here we calculate the square of Dirac equation in a curved space-time according with the Standard Quantum Mechanics. The Dirac equation in a curved space-time is~\cite{Yepez2011,Collas2018}
\begin{equation}\label{eq:Dirac}
    \DD_+\Psi_D=\left(-i\gamma^m\D_m-m_e\right)=0
\end{equation}
where
\begin{equation}\label{eq:D}
   \D_m=D_m -\frac{i}{2}\omega^{pq}_{\quad m} J_{pq}-ieA_m
\end{equation}
$\gamma^m$ are the standard Dirac $\gamma$-matrices in the Weyl spinor representation, viz.
\begin{equation}\label{eq:Diracgamma}
  \gamma^0= \left(
    \begin{matrix}
       0 && -1 \\
       1 && 0
    \end{matrix}
            \right)\hspace{1em}
  {\bm\gamma} = \left(
    \begin{matrix}
       0 && \bm\sigma \\
       -\bm\sigma && 0
    \end{matrix}
           \right)
\end{equation}
and $J_{mn}=-J_{nm}$ is the $4\times 4$ antisymmetric matrix of the generators $J_a=(\bm J,\bm K)$ of rotations ($\bm J$) and boosts ($\bm K$) of the Lorentz group in the Weyl chiral representation $(0,1/2)\oplus(1/2,0)$. The matrix $J_{mn}$ is given by~[see Eq.(\ref{eq:Jmn})]
\begin{equation}\label{eq:Jmnexplicit}
   J_{mn} =\frac{1}{2}
        \left(\begin{matrix}
          0 & -i\alpha_1   & -i\alpha_2 & -i\alpha_3 \\
          i\alpha_1 & 0  & \Sigma_3 & -\Sigma_2 \\
          i\alpha_2 & -\Sigma_3 & 0 & \Sigma_1 \\
          i\alpha_3 & \Sigma_2 & -\Sigma_1 & 0 \\
        \end{matrix}\right)
\end{equation}
where $\bm\alpha = (\alpha_1,\alpha_2,\alpha_3)$ and $\bm\Sigma=(\Sigma_1,\Sigma_2,\Sigma_3)$ are given by
\begin{equation}\label{eq:alpha}
   \bm\alpha = \left(\begin{matrix}
             \bm\sigma & 0  \\
             0 & -\bm\sigma
                \end{matrix}\right)
\end{equation}
\begin{equation}\label{eq:Sigma}
   \bm\Sigma = \left(\begin{matrix}
             \bm\sigma & 0  \\
             0 & \bm\sigma
                \end{matrix}\right)
\end{equation}
with Pauli's matrices $\bm\sigma = (\sigma_1,\sigma_2,\sigma_3)$. Finally, in Eq.~(\ref{eq:Dirac}) $D_m$ is the covariant derivative for the components of a space-time tensor in the local basis~\cite{Yepez2011}. The action of $D_m$ on a Lorentz tensor $T^p_{\ q}$ is given by\footnote{One may also introduce the covariant derivative $D_\mu$ acting on tensors $T^{p\nu}_{\quad q\mu}$ having both Lorentz and absolute indices., but this is not very useful here.}
\begin{equation}\label{eq:Dm}
  D_mT^p_{\ q}=\frac{\partial T^p_{\ q}}{\partial s^m}-\omega^p_{\ rm}T^r_{\ q}+\omega^r_{\ qm}T^p_{\ r}
\end{equation}
The \qo{square} od Dirac's equation is defined as
\begin{eqnarray}\label{eq:Diracsquare}
      \DD_+\DD_-\Psi_D=\DD_-\DD_+\Psi_D&=&
      \left(-i\gamma^m\D_m+m_e\right)\left(-i\gamma^n\D_n-m_e\right)\Psi_D = \nonumber \\
      &=&-\gamma^m\D_m\gamma^n\D_n\Psi_D-m_e^2\Psi_D=0
\end{eqnarray}
where $\DD_\pm$ are the Dirac operators for positive and negative mass, respectively. It is obvious that any solution of the Dirac Eq.~(\ref{eq:Dirac}) is a solution of Eq.~(\ref{eq:Diracsquare}). Solutions of Eq.~(\ref{eq:Diracsquare}) of negative mass can be easily isolated and then discarded as unphysical. To compare Eq.~(\ref{eq:Diracsquare}) with Eq.~(\ref{eq:Dirac2}) of the previous section, we need the following identities
\begin{eqnarray}\label{eq:comm}
   \left[\gamma^n,J^{pq}\right]&=&-i\left(\eta^{np}\gamma^q- \eta^{nq}\gamma^p\right) \label{eq:gammaJcomm} \\
   \left[J_{pq},J_{rs}\right]&=&-i\left(\eta_{ps}J_{qr}+ \eta_{qr}J_{ps}-\eta_{pr}J_{qs}-\eta_{qs}J_{pr}\right)
   \label{eq:JJcomm} \\
   \gamma^m\gamma^n&=&-\eta^{mn}-2iJ^{mn} \label{eq:gammagamma}
\end{eqnarray}
Using Eqs.~(\ref{eq:Dm}) and (\ref{eq:gammaJcomm}) we find
\begin{eqnarray}
  \left[\D_m,\gamma^n\right] &=&
  \left[D_m-\frac{i}{2}\omega^{pq}_{\quad  m}J_{pq},\gamma^n\right]=\left[D_m,\gamma^n\right]-\frac{i}  {2}\omega^{pq}_{\quad m}\left[J_{pq},\gamma^n\right]=  \nonumber \\
  &=&D_m\gamma^n +\frac{i}{2}\omega_{pqm} \left(\gamma^nJ^{pq}-J^{pq}\gamma^n\right)= \nonumber \\
  &=& -\omega^n_{\ qm}\gamma^q+\frac{i}{2}\omega_{pqm}\left[\gamma^n,J^{pq}\right]= \nonumber \\
  &=&-\omega^n_{\  qm}\gamma^q+\frac{1}{2}\left(\eta^{np}\gamma^q-\eta^{nq}\gamma^p\right)= \nonumber \\
  &=&-\omega^n_{\ qm}\gamma^q+\omega^n_{\ qm}\gamma^q=0
\end{eqnarray}
Hence we may write Eq.~(\ref{eq:Diracsquare}) as
\begin{equation}\label{eq:Diracsquare1}
     -\gamma^{m}\gamma^{n}\D_m\D_n\Psi_D-m_e^2\Psi_D=0
\end{equation}
Then, using Eq.~(\ref{eq:gammagamma}) and the antisymmetry of $J^{mn}$ we find
\begin{equation}\label{eq:Diracsquare2}
   \eta^{mn}\D_m\D_n\Psi_D+2iJ^{mn}\left[\D_m,\D_n\right]\Psi_D -m_e^2\Psi_D=0
\end{equation}
A straightforward calculation based on Eqs.~(\ref{eq:Dm}) and (\ref{eq:JJcomm}) yields
\begin{equation}\label{eq:Dcomm}
  \left[\D_m,\D_n\right] = -\frac{i}{2}\left(e F_{mn}-R^{pq}_{\quad mn}J_{pq}\right)
\end{equation}
where $R^{pq}_{\quad mn}$ is the space-time Riemann tensor.
Inserting Eq.~(\ref{eq:Dcomm}) into the Dirac Eq.~(\ref{eq:Dirac2}) yields
\begin{equation}\label{eq:Diracsquare3}
   \left(\eta^{mn}\D_m\D_n + e F_{mn}J^{mn}-m_e^2- R_{pqmn}J^{mn}J^{pq}\right)\Psi_D=0
\end{equation}
We may further simplify the last equation by using the symmetry $R_{pqmn}=R_{mnpq}$ and the anticommutator for the Lorentz generators in the $(1/2,0)\oplus(1/2,0)$ representation
\begin{equation}\label{eq:JJanticomm}
  \{J^{pq},J^{mn}\}=\frac{1}{2}\left(\eta^{pm}\eta^{qn}- \eta^{pn}\eta^{qm}-i\epsilon^{pqmn}\gamma^5\right)
\end{equation}
where $\gamma^5=i\gamma^0\gamma^1\gamma^2\gamma^3=\left(\begin{smallmatrix} 1 & 0 \\ 0 & -1\end{smallmatrix}\right)$. Inserting Eq.~(\ref{eq:JJanticomm}) into Eq.~(\ref{eq:Diracsquare3}) we find
\begin{equation}\label{eq:Diracsquare4}
   \left[\eta^{mn}\D_m\D_n + e F_{mn}J^{mn}-m_e^2- \frac{1}{2}\left(R_4+i\ ^*\!R_4\gamma^5\right)\right]\Psi_D=0
\end{equation}
where $R_4$ is the space-time Riemann scalar curvature and $^*\!R_4$ is its right dual, which actually is zero~\cite{Shen2004}. Finally, after passing to the absolute components and using the identity $F_{mn}J^{mn}={\bm H}\cdot{\bm\Sigma}-i{\bm E}\cdot{\bm \alpha}$, the square of Dirac's equation in the standard Quantum Mechanics is
\begin{eqnarray}
   &&g^{\mu\nu}\left[\left(-i\nabla_\mu-\frac{1}{2}\omega^{pq}_{\quad \mu}(x) J_{pq}-e A_\mu(x)\right)\left(-i\nabla_\nu-\frac{1}{2} \omega^{pq}_{\quad \nu}(x) J_{pq}-e A_\nu(x)\right)\right]\Psi_D - \nonumber\\
   &{}-& e \left(\bm H(x)\cdot\bm\Sigma-i\bm E(x)\cdot\bm\alpha\right)\Psi_D+\left(m_e^2+\frac{1}{2}R_4\right) \Psi_D=0
\end{eqnarray}
with $\D_m$ given by Eq.~(\ref{eq:Dm}). We note the presence of the space-time curvature as a scalar potential acting on the electron. The same happens in our Eq.~(\ref{eq:Dirac2}) with the difference that the coupling constant is $\gamma^2\simeq 1/4$ instead of the quantum result $1/2$.
%
%
%
%

\end{document}